\documentclass{aa}  

\usepackage{graphicx}
\usepackage{txfonts}
\usepackage{placeins}           
                                
\begin{document}
\nolinenumbers
   \title{Inferring physical parameters of solar filaments from simultaneous longitudinal and transverse oscillations}

   \author{Upasna Baweja\inst{1,2}, 
        Vaibhav Pant\inst{1}, Iñigo Arregui \inst{3,4}
        \and M. Saleem Khan \inst{2}
        }

   \institute{Aryabhatta Research Institute of Observational Sciences, 263001, Nainital, India\\
        \and Department of Applied Physics, Mahatma Jyotiba Phule Rohilkhand University, Bareilly- 243006, Uttar Pradesh, India\\
    \email{upasnabaweja.ub@gmail.com,vaibhavpant55@gmail.com}
        \and Instituto de Astrof\'{\i}sica de Canarias, E-38205 La Laguna, Tenerife, Spain\\
        \and Departamento de Astrof\'{\i}sica, Universidad de La Laguna, E-38206 La Laguna, Tenerife, Spain\\ }
 
 
  \abstract
   {Different modes of oscillations are frequently observed in solar prominences/filaments, and prominence seismology helps estimate important physical parameters like the magnetic field strength. Although the simultaneous detection of longitudinal and transverse oscillations in the same filament is not common, such rare observations provide a unique opportunity to constrain the physical parameters of interest.}
   {In this study, we aim to estimate the physical parameters of prominences undergoing simultaneous longitudinal and transverse oscillations.} 
   {We apply Bayesian seismology techniques to observations of longitudinal and transverse filament oscillations to infer the magnetic field strength, the length, and the number of twists in the flux tube holding the prominence plasma. We first use the observations of longitudinal oscillations and the pendulum model to infer the posterior probability density for the magnetic field strength. The obtained marginal posterior of the magnetic field, combined with the observations of the transverse oscillations, is then used to estimate the probable values of the length of the magnetic flux tube that supports the filament material using Bayesian inference. This estimated length is used to compute the number of twists in the flux tube.}
   {For the prominences under study, we find that the length of the flux tubes supporting the quiescent prominences can be very large (from 100 to 1000 Mm) and the number of twists in the flux tube are not more than three.}
   {Our results demonstrate that Bayesian analysis offers valuable methods to perform parameter inference in the context of prominence seismology.}

\titlerunning{Inferring physical parameters of solar filaments}
\authorrunning{Baweja et al.}
   \maketitle
\nolinenumbers
\section{Introduction}
Understanding the internal structure, dynamics, and energetics of solar prominences is challenging. However, the presence of oscillations in the filaments, detected using H$\alpha$ spectrograms \citep{dyson1930solar} and extreme ultraviolet (EUV) observations \citep{2012ApJ...760L..10L,2013ApJ...773..166L} are a useful source of information. Depending upon the velocity amplitudes, prominence oscillations are classified into two categories \citep{2002SoPh..206...45O,2018LRSP...15....3A}. Large amplitude oscillations in which the entire prominence or the major part of it experiences oscillations with velocity amplitude greater than 10 km s$^{-1}$ \citep{2004ApJ...608.1124O,2009SSRv..149..283T,2022ApJ...940...34K, 2023A&A...672A..15J}. On the other hand, small amplitude oscillations affect the localised regions of the prominence and have velocity amplitudes of a few (2-3) km s$^{-1}$ \citep{1991SoPh..134..275Y, 1991SoPh..132...63Y,engvold2000encyclopedia,oliver2001encyclopedia,oliver2002oscillations,2009A&A...499..595N,2011ApJ...726..102S}. 

Prominences support both longitudinal and transverse oscillations. If the periodic motion of the plasma is along the magnetic field, the oscillations are called longitudinal oscillations. First observations of such longitudinal oscillations were reported using H$\alpha$ observations made at the Big Bear Solar Observatory \citep{2003ApJ...584L.103J,2006SoPh..236...97J}. Longitudinal oscillations are triggered when a small energetic event, such as a sub-flare, flare, or small jet, occurs near the footpoint of the filament \citep{2003ApJ...584L.103J, 2006SoPh..236...97J,2014ApJ...785...79L,2017ApJ...851...47Z, 2021ApJ...923...74D}. Not only triggering but sometimes, nearby flares can also enhance the longitudinal oscillations \citep{2020A&A...635A.132Z}. Additionally, the merging of two filaments can also trigger large amplitude longitudinal oscillations \citep{2017ApJ...850..143L}. Initially, magnetic tension \citep{2003ApJ...584L.103J,2006SoPh..236...97J} and/or magnetic pressure gradient \citep{2007A&A...471..295V} were considered to be the restoring force of such oscillations. Later, \cite{2012ApJ...750L...1L,2012ApJ...757...98L,2022A&A...660A..54L} provided a theoretical model known as the \textit{pendulum model} to explain these longitudinal oscillations. 

Apart from these longitudinal oscillations, motions perpendicular to the magnetic field direction are also observed. Depending on the direction of oscillations, they can be either vertical or horizontal motions of the filaments and are collectively known as transverse oscillations of the filaments \citep{1966AJ.....71..197R,1966ZA.....63...78H,1969SoPh....6...72K,2009ApJ...704..870L,2011A&A...531A..53H,  2012ApJ...753...52L, 2012ApJ...761..103G,2015RAA....15.1713P,2022AdSpR..70.1592D,2023ApJ...959...71D, 2024arXiv240115858Z}.

A few studies have reported simultaneous longitudinal and transverse oscillations in the same prominence/prominence threads \citep{2007Sci...318.1577O, 2008ApJ...685..629G, 2016SoPh..291.3303P,2023MNRAS.520.3080T}. \cite{2016SoPh..291.3303P} reported both longitudinal and transverse oscillations in an active region filament triggered after being hit by a shock wave. \cite{2017ApJ...851...47Z} reported longitudinal and transverse oscillations in the same quiescent prominence triggered by a jet. \cite{2020AA...633A..12M, 2021ApJ...923...74D} also reported the simultaneous presence of longitudinal and transverse oscillations in quiescent filaments. Recently, both wave modes have been reported to be excited in a filament triggered by EUV waves \citep{2025MNRAS.542.1308P}. Additionally, the three-dimensional motions of the filaments are also probed using observations of horizontal and vertical transverse oscillations \citep{2006A&A...449L..17I, 2023ApJ...959...71D}. 

Combining these observations with magnetohydrodynamic (MHD) wave theory can help estimate the physical parameters, such as magnetic field strength, using seismological tools \citep{2018LRSP...15....3A}. Prominence seismology was first suggested by \cite{1995ASSL..199.....T} after the successful application of seismology in coronal loops \citep{1970PASJ...22..341U,1984ApJ...279..857R}. Recently, \cite{2016SoPh..291.3303P,2020AA...633A..12M,2021ApJ...923...74D,2023MNRAS.520.3080T,2025MNRAS.542.1308P} have employed the so-called pendulum model to estimate the filament's magnetic field and the radius of curvature of the dip, from observations of their longitudinal oscillations. \cite{2020AA...633A..12M} further used this magnetic field to estimate the length of the magnetic field lines supporting the prominence material. These previous studies provide either point estimates and/or possible ranges of variation for the parameters of interest.

The use of Bayesian analysis in prominence seismology was first suggested by \cite{2014IAUS..300..393A} and successfully applied to infer the magnetic field strength and transverse density length-scales in prominence threads by \cite{2019A&A...622A..88M}. The main advantage of Bayesian methods for seismology inversions is the consistent treatment of information that is incomplete and uncertain, which makes impossible to obtain a unique solution \citep{2019A&A...622A..44A}, so this has to be formulated in the form of probability density distributions \citep{2018AdSpR..61..655A,2022FrASS...926947A}. The methods also propagate correctly the errors in observed data into uncertainty in the inferred parameters \citep{2011ApJ...740...44A}.

This paper aims to investigate the application of Bayesian inference in analysing simultaneous longitudinal and transverse oscillations within the same prominence to estimate the associated physical parameters. To achieve this, we adopt the methodology similar to \cite{2020AA...633A..12M}. Firstly, the magnetic field strength supporting the prominence thread is inferred, which is subsequently used to determine the length of the flux tube, which is further used to estimate the twist number. However, in contrast to their study, we explored the whole parameter space within the Bayesian framework. Thus, by considering the appropriate prior distributions, we obtain the posterior distribution of both length and the magnetic field strength holding the quiescent prominence flux tube. First, a brief description of Bayesian methodology is presented in Section~\ref{sec:Bayesian_statistics}. Then, using the analysis of longitudinal oscillations in \cite{2020AA...633A..12M, 2025MNRAS.542.1308P}, the probable values of magnetic fields are obtained in Section~\ref{sec:longitudinal_oscillations}. Combining these probable values of the magnetic field and analysis of the transverse oscillations, the length of the field lines supporting the prominence material is determined in Section~\ref{sec:transverse_oscillations}. The number of twists associated with these flux tubes is also estimated in Section \ref{sec:twist_number} after computing the radius of curvature of these field lines. Section~\ref{sec:Summary and conclusions} will briefly summarise the work.

\section{Methodology: Bayesian statistics}\label{sec:Bayesian_statistics}
Given a model $M$, with parameter vector $\theta$, proposed to explain observed data $d$, Bayesian parameter inference relies on the use of  Bayes Theorem 
\begin{equation}
\label{equation:bayes formula}
p(\theta|d,M) = \frac{p(d|\theta, M)p(\theta|M)}{\int p(d|\theta, M)p(\theta|M)d\theta}.   
\end{equation}
Here, the posterior probability distribution $p(\theta|d, M)$ is obtained from the combination of the likelihood function $p(d|\theta, M)$ and the prior probability density $p(\theta|M)$ and encompasses all the information that can be gathered from the assumed model and the observed data.

For multiparameter models, $\theta = \{\theta_1,...,\theta_i,...,\theta_N\}$, the probability of a particular parameter of interest can be obtained by marginalising the posterior probability distribution with respect to all the other parameters as follows:
\begin{equation}\label{basic_marginal}
    p(\theta_i|d, M) = \int p(\theta|d,M)d\theta_1\,...\,d\theta_{i-1},d\theta_{i+1}\,...\,d\theta_N.   
\end{equation}\par
Thus, for parameter estimation using Bayesian inference, two pieces of information are required: the prior information and the likelihood function. In this study, all the parameters are considered to be independent; thus, the global prior is the product of the individual priors associated with each parameter. Three types of individual priors are considered here. Firstly, we considered that each parameter lies in a given plausible range, where all the values are equally probable, then a uniform prior of the form
\begin{equation}\label{uniform_prior_theta}
    \mathcal{U}(\theta_i; \theta_{i_{\rm min}},\theta_{i_{\rm max}})=\begin{cases}
    \frac{1}{\theta_{i_{\mathrm{max}}} - \theta_{i_{\mathrm{min}}}} & \text{$\theta_{i_{\mathrm{min}}} \le \theta_i\le \theta_{i_{\mathrm{max}}}$} \\ 
    0 & \text{otherwise},
    \end{cases}
\end{equation}
is adopted.  Second, when more specific information about the parameter is available from the observations, the gamma and Gaussian distributions are used as priors. The gamma distribution is considered positive and has no upper bound. A parametric form of the gamma distribution can be expressed as:
\begin{equation}\label{gamma_prior_theta}
    \gamma(\theta_i;\alpha,\beta) =\begin{cases}
    \frac{\beta^\alpha \theta_i^{\alpha-1}e^{-\beta \theta_i}}{\Gamma(\alpha)} & \text{$\theta_{i_{\textit{min}}} \le \theta_i\le \theta_{i_{\textit{max}}}$} \\ 
    0 & \text{otherwise}.
    \end{cases}
\end{equation} 
Here, $\alpha$ and $\beta$ correspond to the shape and rate parameters. The distribution's mean and coefficient of variance are given by \(\alpha/\beta\) and \(\alpha/\beta^2\), respectively. Alternatively, for a Gaussian prior, the mean ($\mu_{\theta_i}$) and the uncertainty ($\sigma_{\theta_i}$) associated with the parameter can be used, and its distribution is of the form
\begin{equation}\label{equation:Gaussian_prior}
    \mathcal{G}(\theta_i; \mu_{\theta_i},\sigma_{\theta_i}) = \frac{1}{\sqrt{2\pi}\sigma_{\theta_i}}  \exp{\left[\frac{-(\theta_i - \mu_{\theta_i})^2}{2\sigma_{\theta_i}^2}\right]}.
\end{equation}

For the likelihood functions, Gaussian profiles are applied according to the normal error assumption, which can be expressed as:
\begin{equation}\label{equation:likelihood_theta}
    p(d|\theta, M) = \frac{1}{\sqrt{2\pi}\sigma_{d}}  \exp{\left[\frac{-(d - d_{M}(\theta))^2}{2\sigma_{d}^2}\right]}.
\end{equation}
The exponential of the Equation~\ref{equation:likelihood_theta} explicitly contains its dependence on the model \textit{M}, which is assumed to be true. Notably, it does not convey the likelihood of various data occurrences but serves to quantify the disparity between model predictions and observed data relative to the uncertainty in the data. Consequently, it assigns varying likelihood levels to alternative parameter combinations. 

Calculating posteriors and marginal probability distributions necessitates solving integrals within the parameter space. When dealing with a low-dimensional problem, for instance, in this work, direct numerical integration over a grid of points is viable, and the sampling of the posterior using Markov Chain Monte Carlo (MCMC) methods serves as a practical alternative, applicable not only in low-dimensional but also in high-dimensional spaces. We follow the same approach as \cite{2017ApJ...846...89M}, \cite{2019A&A...622A..88M}, \cite{2024ApJ...963...69B}, and \cite{2025ApJ...991..208Z} and compute the required posteriors using both direct integration and MCMC sampling, ensuring that the results are consistent. \par

\section{Analysis and results}
In our analysis, we employed Bayesian methods to estimate the prominence physical parameters of interest, combining theoretical models and observations of simultaneous transversely and longitudinally oscillating quiescent prominences studied in \cite{2020AA...633A..12M}, and \cite{2025MNRAS.542.1308P}. First, we use the period of longitudinal oscillations to constrain the magnetic field strength. Then, this information is used together with the period of transverse oscillations to infer the length of the magnetic flux tube holding the prominence/filament plasma. Finally, we estimate the twist number.

\subsection{Inference of magnetic field strength from longitudinal oscillations}\label{sec:longitudinal_oscillations}
To estimate the magnetic field strength from the longitudinal oscillations, we adopt the pendulum model \citep{2012ApJ...750L...1L}, in which the magnetic tension balances the projected gravity in the flux tube dip, such that
\begin{equation}
\label{pendulum model}
\frac{B^2}{r} - mn_eg  \ge 0.
\end{equation}
Here $B$ is the magnetic field strength at the bottom of the dip, $r$ is the radius of curvature of the dips containing the cool prominence plasma, \(m = 1.27 m_p\) (with $m_p$ the proton mass), the mean particle mass \citep{2005psci.book.....A}, and $n_e$ the electron number density in the thread. The model has been verified by multiwavelength analysis and numerical simulations of an active region prominence by \cite{2012A&A...542A..52Z}. Recently, \cite{2022A&A...660A..54L} have extended the model to consider nonuniform gravity and non-circular dips.

For the plasma to oscillate longitudinally in the prominences, the surface gravity ($g_0$) of the Sun is the main restoring force and thus, the angular frequency ($\omega$) of these longitudinal oscillations is given by:
\begin{equation}
\label{equation:angular_freq}
\centering
\begin{aligned}
\omega = \frac{2 \pi}{p_l} = \sqrt{\frac{g_0}{r}}.
\end{aligned}
\end{equation}
Here, $p_l$ denotes the period of the longitudinal oscillations. Combining this equation with the pendulum model (Equation~\ref{pendulum model}) and substituting the mass $m$ and $g_0$ = 274 m s$^{-1}$, a lower limit for the magnetic field strength in the prominence can be obtained as
\begin{equation}
\label{equation:longitudinal_oscillations}
\centering
\begin{aligned}
B[{\rm G}] = 26\sqrt{\frac{n_{\rm e}}{10^{11} {\rm cm}^{-3}}}\,p_l\,[{\rm hr}],
\end{aligned}
\end{equation}
provided the periodicity $p_l$ of the longitudinal oscillations is known. This equation can be further simplified as: 
\begin{equation}
\label{equation:Model_for_LO}
\centering
\begin{aligned}
p_l=\frac{B}{26}\sqrt{\frac{10^{11}}{n_{e}}},
\end{aligned}
\end{equation}

In what follows, we use $p_l$ and $P_l$ to denote theoretical and observational longitudinal periods, respectively.
Observationally, \cite{2018ApJS..236...35L} analysed longitudinal oscillations in 196 filaments and found the mean periodicity to be $P_l=58\pm15$ minutes. They assumed the typical values of electron number density (\(n_e = 10^{10}-10^{11} {\rm cm}^{-3}\), \citealt{2014ApJ...785...79L}), in the pendulum model (Equation \ref{pendulum model}) and obtained the average minimum value of the magnetic field ($\approx$16 G) and radius of curvature of the magnetic dips in the filaments ($\approx$89 Mm). 

\begin{figure}[t!]
    \centering
    \includegraphics[width=0.45\textwidth]{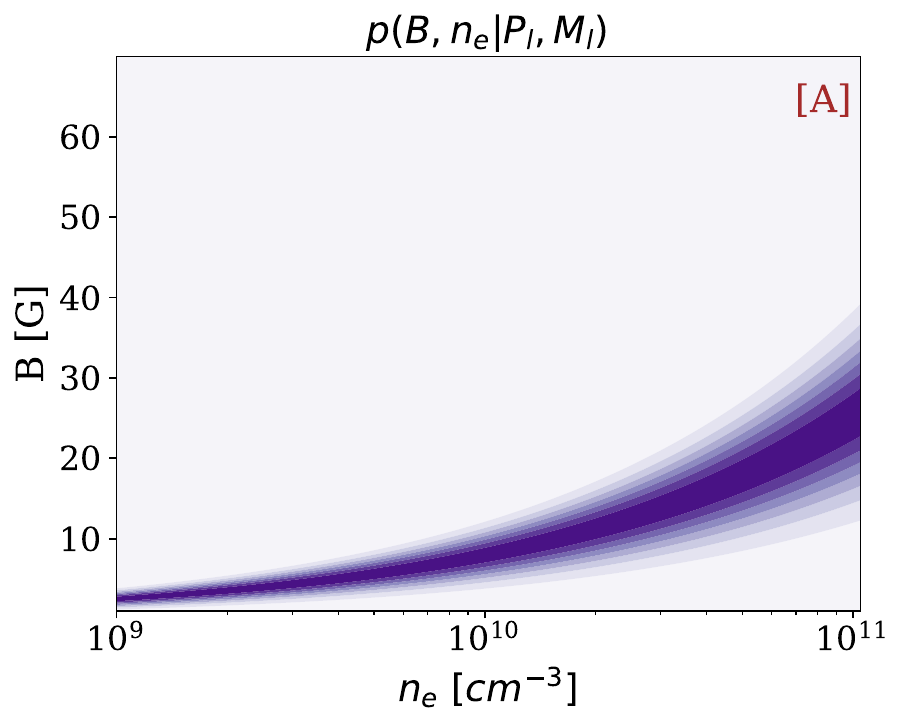}
    \includegraphics[width=0.45\textwidth]{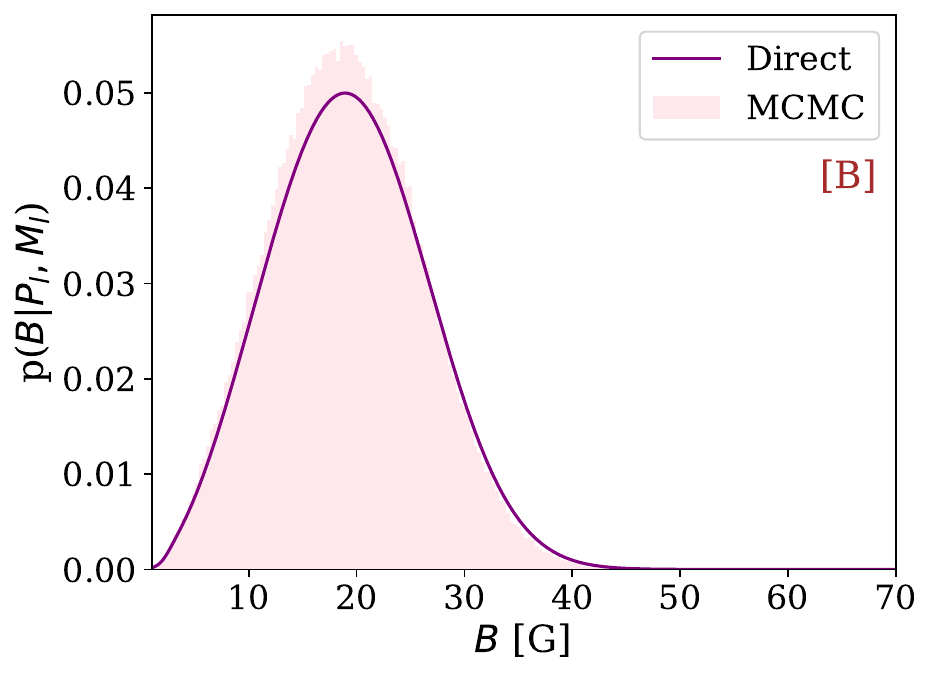}
    \caption{Joint posterior distribution of magnetic field strength and electron density and marginal probability distribution of magnetic field. Panel (A) shows the joint probability distribution of magnetic field strength and electron number density inferred for the average longitudinal oscillation period $P_l = 58 \pm 15$ min reported by \cite{2018ApJS..236...35L}, assuming uniform priors on electron density, $\mathcal{U}(n_{\rm e},[\mathrm{cm}^{-3}]; 10^{9}, 10^{11})$, and magnetic field strength, $\mathcal{U}(B,[\mathrm{G}]; 1, 70)$, evaluated on a two-dimensional grid with $N_{n_{\rm e}} = 990$ and $N_B = 1380$ points. Panel (B) presents the marginal probability distribution of the magnetic field obtained from direct numerical integration (solid line) and from the \textit{emcee} MCMC sampling (pink histograms), using a two-dimensional parameter space with approximately 15 walkers, 50 000 steps per walker, and a burn-in phase discarding the first 20\% of the iterations.}
    \label{figure:longitudinal_general}
\end{figure}

In our Bayesian analysis, the Equation~\ref{equation:Model_for_LO} is considered as our model $M_l$ to obtain the probable range of the minimum magnetic field strength from observed periods of longitudinal oscillations $P_l$. This model is dependent on two parameters, namely $\theta = \{n_e, B\}$. Following \cite{2010SSRv..151..243L}, we adopt uniform priors $\mathcal{U}(n_{\rm e}$ [cm$^{-3}$]; 10$^9$, 10$^{11}$) and $\mathcal{U}(B [G]; 1, 70)$ for the electron density and the magnetic field strength. We forward model the oscillation period $p_l$ by applying model $M_l$ (described by Equation~\ref{equation:Model_for_LO}) to explore parameter values based on specified priors. These theoretical predictions $p_l$ are compared to the observed period $P_{l}$ of longitudinal oscillations (58$ \pm $15 minutes). The relative merit of each combination is assessed by adopting a Gaussian likelihood function of the form described in Equation \ref{equation:likelihood_theta}. The joint posterior distribution of $B$ and $n_e$ obtained through application of Bayes theorem, Equation~\ref{equation:bayes formula} is: 
\begin{equation}\label{equation:posterior}
    p(B,n_e|P_l,M_l) = \frac{p(P_l|B,n_e, M_l)p(B)p(n_e)}{\int \int p(P_l|B,n_e,M_l)p(B)p(n_e)dB dn_e},
\end{equation}
where $p(P_l|B,n_e, M_l)$ is the likelihood function and $p(B)$, $p(n_e)$ are the prior probability distribution functions of $B$ and $n_e$, respectively. The joint probability distribution ($p(B,n_e|P_l,M_l)$) is shown in Figure \ref{figure:longitudinal_general} (A) and the marginal distribution of $B$ obtained by marginalising the joint probability distribution in Figure \ref{figure:longitudinal_general} (B). The results indicate that the magnetic field strength can be properly inferred, even if the electron density is largely unknown. The maximum a posteriori estimate for $B\approx$19 G is 3 G larger than the point-estimate in \cite{2018ApJS..236...35L} ($\approx$16 G). Our posterior contains that point-estimate and, in addition, offers the full probability over all values of $B$ in the considered range. This enables uncertainty quantification and making direct probability statements, such as the probability that the magnetic field strength is between 5.5 and 33 G is 95\%, for example.

The results obtained by direct integration over a grid of points are further validated with the use of  Markov chain Monte Carlo (MCMC) sampling for the posterior distribution using the \textit{emcee} algorithm \citep{2013PASP..125..306F}. The methodology and application of the \textit{emcee} algorithm is explained in \cite{2017ApJ...846...89M,2019A&A...625A..35A,2024ApJ...963...69B}. In Figure \ref{figure:longitudinal_general} (B), the pink histogram represents the results from MCMC, and the solid line represents the direct integration result. Both approaches lead to very similar posteriors, which gives us confidence on the obtained results.

The procedure is next applied to the two observations with simultaneous longitudinal and transverse oscillations reported by \cite{2020AA...633A..12M} and \cite{2025MNRAS.542.1308P}.
The measured periods ($P_l$) for each case are shown in the second column of Table~\ref{table:longitudinal_parameters}.
The uncertainties in $P_l$ are not given in \cite{2025MNRAS.542.1308P}; thus, 10$\%$ uncertainty is assumed \citep{2019A&A...622A..88M}. 
\begin{table}
    \centering
    \caption{Longitudinal oscillation parameters and inferred magnetic field: References, longitudinal oscillation period ($P_l$), magnetic field strengths reported in those references ($B$), and the most probable magnetic field inferred using uniform ($B_u$), Gaussian ($B_G$), and gamma ($B_\gamma$) priors, with uncertainties quoted at the 68\% credible interval.}
    \label{table:longitudinal_parameters}
    \setlength{\tabcolsep}{3pt} 
    \begin{tabular}{@{}c c c c c c@{}}
    \hline
    Reference& \shortstack{$P_l$\\(min)}&\shortstack{$B$\\(G)}&\shortstack{$B_u$\\(G)}&\shortstack{$B_G$\\(G)}&\shortstack{$B_\gamma$\\(G)}\\
    \hline
    (1) & 79.8 & 22.6 & 31.1$^{+3.6}_{-9.0}$ & 29.7$^{+3.9}_{-8.2}$ & 28.3$^{+4.5}_{-9.1}$ \\
    (2) & 18.31 & 5.1 & 6.8$^{+1.2}_{-2.1}$ & 5.0$^{+0.5}_{-0.5}$ & 5.1$^{+0.5}_{-0.5}$ \\
    \hline
    \end{tabular}
    \tablebib{(1)~\citet{2020AA...633A..12M}; (2) \citet{2025MNRAS.542.1308P}.}
\end{table}

\begin{figure}[t]
    \centering
    \includegraphics[width=0.45\textwidth]{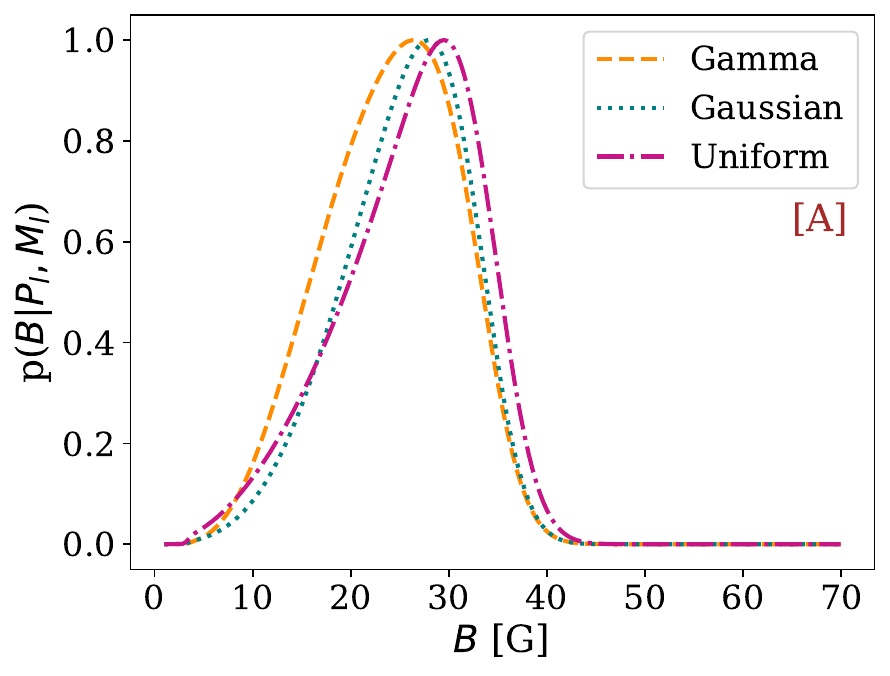}
    \includegraphics[width=0.45\textwidth]{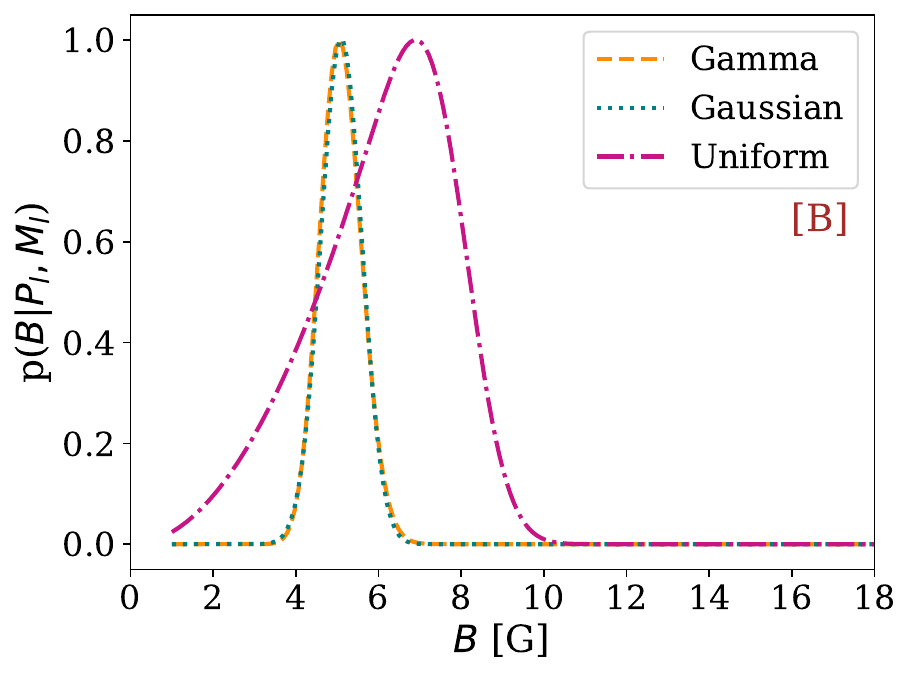}
    \caption{Marginal posterior distributions of the magnetic field. Marginal probability distributions inferred using uniform (violet dot–dashed), gamma (orange dashed), and Gaussian (teal dotted) magnetic-field priors from longitudinal oscillations in panels (A) \cite{2020AA...633A..12M} and (B) \cite{2025MNRAS.542.1308P}, with all distributions normalised to their maximum values. The direct solutions assume uniform priors on the electron number density, $\mathcal{U}(n_{\rm e},[\mathrm{cm}^{-3}]; 10^{9}, 10^{11})$. For the magnetic field, the priors $\mathcal{U}(B,[\mathrm{G}]; 1, 70)$, $\gamma(B; 3.6, 0.2)$, and $\mathcal{G}(B,[\mathrm{G}]; 22.6, 11.9)$ are used for panel (A), while $\mathcal{U}(B,[\mathrm{G}]; 1, 70)$, $\mathcal{G}(B,[\mathrm{G}]; 5.1, 0.5)$, and $\gamma(B; 100, 20)$ are adopted for panel (B).}
    \label{figure:marginal_longitudinal}
\end{figure}

Additionally, based on the estimated values of $B=22.6\pm11.9$ G and $B=5.1\pm0.5$ G in \cite{2020AA...633A..12M} and \cite{2025MNRAS.542.1308P}, respectively, we can now construct more informed priors described by Equations \ref{gamma_prior_theta}, and \ref{equation:Gaussian_prior} along with the uniform prior (Equation \ref{uniform_prior_theta}). The marginal posterior distributions of $B$ obtained for different priors and both studies are shown in Figure \ref{figure:marginal_longitudinal}. The most probable values of $B$ obtained from each prior and case are reported in Table \ref{table:longitudinal_parameters}.

From Figure \ref{figure:marginal_longitudinal}, we find that for both observations and all prior types, the marginal posterior distributions for the magnetic field strength can be well-constrained.  There are some differences in their shape and maximum a posteriori estimates, depending on the prior used for the magnetic field strength. For the observation by \cite{2020AA...633A..12M} (Figure~\ref{figure:marginal_longitudinal} (A)), the three posteriors are very similar and span almost identical ranges. For the observation by \cite{2025MNRAS.542.1308P} (Figure~\ref{figure:marginal_longitudinal} (B)), the posterior for the uniform prior is the least constrained, while the assumption of Gaussian and $\gamma$ priors in $B$ leads to almost identical posterior distributions.

Since the shape of the posterior depends on both the likelihood and the prior distribution, when a uniform prior is used, the posterior shape is primarily governed by the likelihood function. The likelihood depends on the mean and standard deviation of the measured periodicity of the longitudinal oscillations. The relative uncertainty in the periodicity reported by \cite{2018ApJS..236...35L} is about 25\%, whereas it is around 10\% and less than 1\% for \cite{2025MNRAS.542.1308P} and \cite{2020AA...633A..12M}, respectively. The smaller the uncertainty in the period, the narrower the joint probability $p(B,n_e|P_l,M_l)$ is. This explains why the posterior in Figure~\ref{figure:longitudinal_general} resembles a Gaussian distribution, while those in Figure~\ref{figure:marginal_longitudinal} (A) and (B) for the case of uniform prior have a sharper drop-off rate to the right of the maximum probability.

Furthermore, the relative uncertainties in the magnetic field values reported by \cite{2025MNRAS.542.1308P} and \cite{2020AA...633A..12M} are 9.8\% and 52.6\%, respectively. Therefore, for \cite{2025MNRAS.542.1308P}, the priors on $B$ derived from both the gamma and Gaussian distributions are nearly similar, whereas for \cite{2020AA...633A..12M}, they differ significantly and thus influence the posterior shape more strongly. To verify this, we varied the uncertainty in $B$ and found that as the uncertainty increases, the priors derived from the gamma and Gaussian distributions become increasingly different, leading to noticeable changes in the posterior. When the uncertainty in $B$ is sufficiently large, the posterior distributions in Figure~\ref{figure:marginal_longitudinal} (A) and (B) become more similar.

\subsection{Inference of the length of the magnetic flux tube from transverse oscillations}\label{sec:transverse_oscillations}

After obtaining the probable magnetic field values, the next target is to estimate the length of the magnetic flux tube containing the prominence plasma from the seismology of the transverse oscillations. For this, we considered the prominence model for non-flowing prominence threads given by \cite{2002ApJ...580..550D} and \cite{2005SoPh..229...79D}. The equilibrium configuration of the filament is similar to the one shown in Figure 7 of \cite{2020AA...633A..12M}. The length of the prominence thread is $2W$, having density $\rho_p$, embedded in a magnetic flux tube of length $2L$. The density of the evacuated parts of the flux tube is $\rho_e$ such that it is larger than the density of the ambient corona $\rho_c$. The piecewise density along the flux tube is given by Equation 5 of \cite{2020AA...633A..12M} and is similar to the one considered in \cite{2005SoPh..229...79D}. \cite{2002ApJ...580..550D} and \cite{2005SoPh..229...79D} studied the normal modes of non-flowing filament threads ($v_0=0$), whereas \cite{Terradas_2008} did a similar seismological analysis for flowing threads ($v_0\neq0$). The results of \cite{Terradas_2008} indicate that the effect of flow is negligible, supporting the assumption of non-flowing threads and thereby validating the use of the model presented by \cite{2005SoPh..229...79D} in this analysis. Additionally, transverse oscillations are observed to occur in the same thread that undergoes longitudinal oscillations in \cite{2020AA...633A..12M}. In \cite{2025MNRAS.542.1308P}, the prominence is undergoing both longitudinal and transverse oscillations simultaneously and from the slit positions (S3 and S4 in Figure 5 of \citealt{2025MNRAS.542.1308P}), it is assumed that the same thread is undergoing simultaneous oscillations. Thus, the assumptions by \cite{2005SoPh..229...79D} are valid in these two studies. Further, in \cite{2005SoPh..229...79D}, the system supports Alfv\'en and fast waves; however, the observed transverse oscillations are the result of the displacement of the cylindrical axis, which is caused by the kink-fast mode. Assuming the thin tube approximation, \cite{2005SoPh..229...79D} provided a simple dispersion relation given as:
\begin{equation}
\begin{aligned}
\tan\Big(\Omega(1-l)\sqrt{\frac{1+e}{2}}\Big)-\sqrt{\frac{1+e}{1+c}}\cot\Big(\Omega l \sqrt{\frac{1+c}{2}}\Big)=0
\end{aligned}
\label{equation:transcedental_equation}
\end{equation}
Here, $\Omega=\omega L/v_{Ac}$, $e=\rho_{e}/\rho_{c}$, $c=\rho_{p}/\rho_{c}$, $l=W/L$, $v_{Ac}$ is the coronal Alfvén speed, $\omega$ is the frequency of oscillations, and L is the half length of flux tube.
\begin{figure*}
    \centering
    \includegraphics[width=0.45\textwidth]{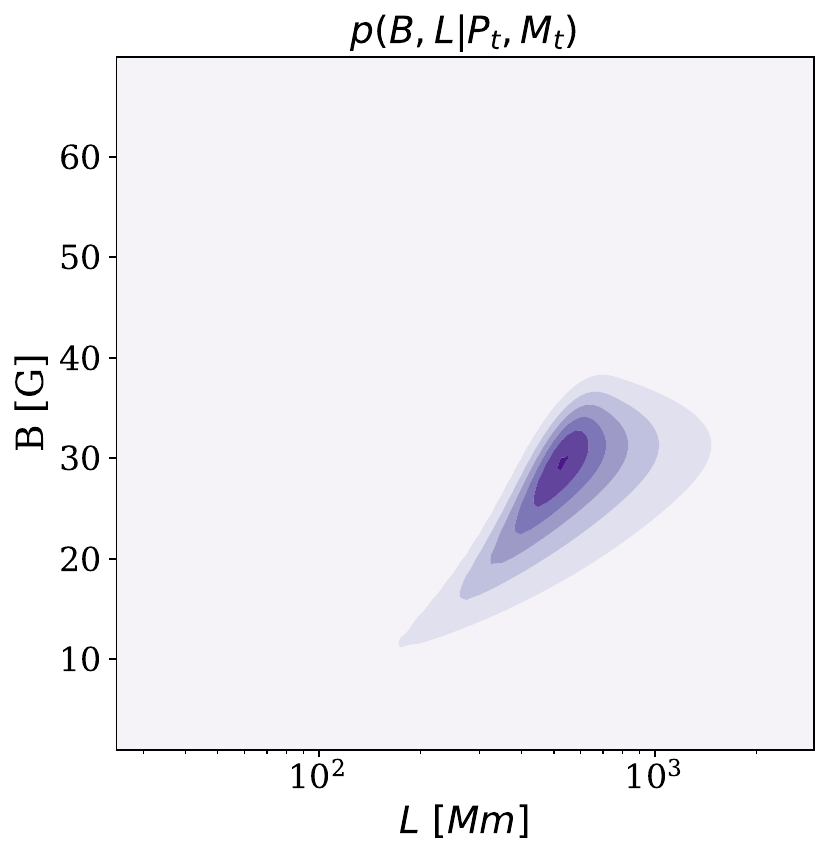}
    \includegraphics[width=0.45\textwidth]{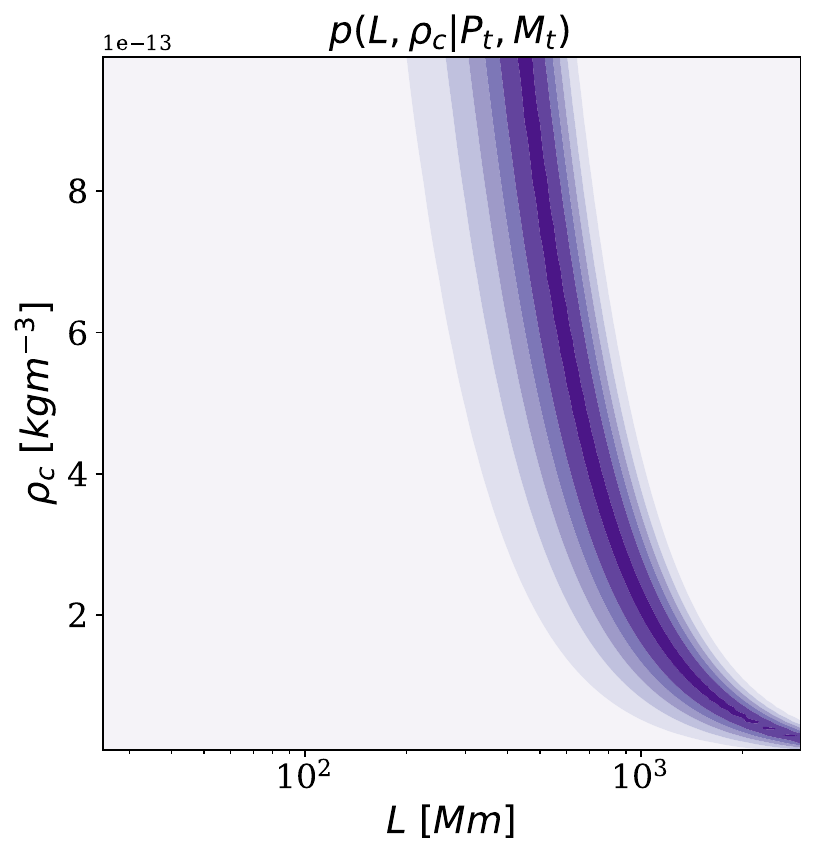}
    \includegraphics[width=0.45\textwidth]{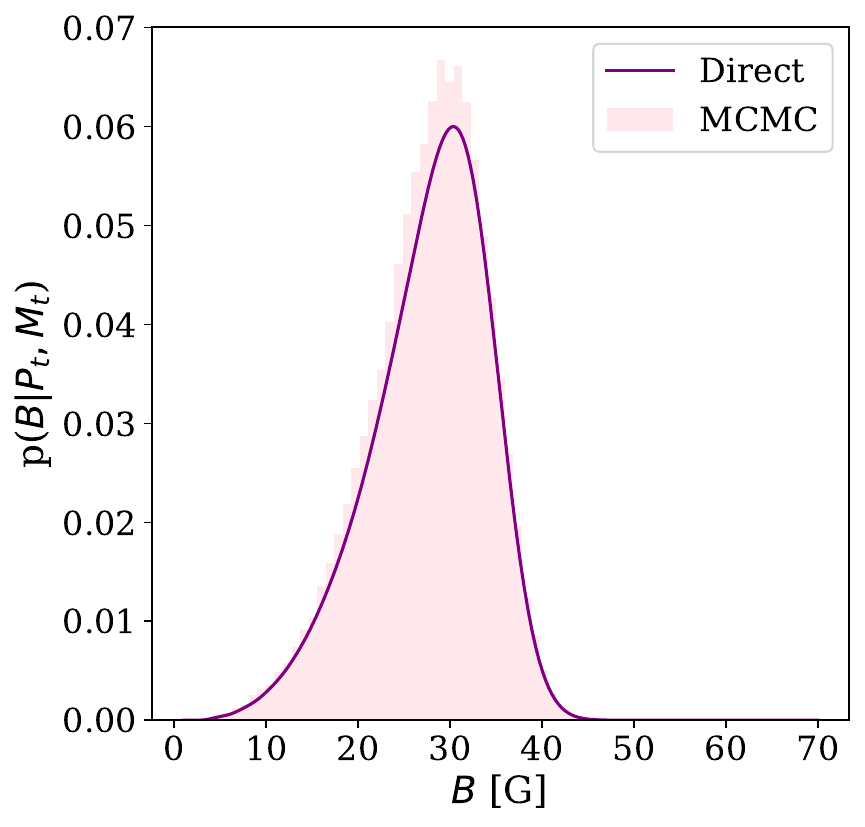}
    \includegraphics[width=0.45\textwidth]{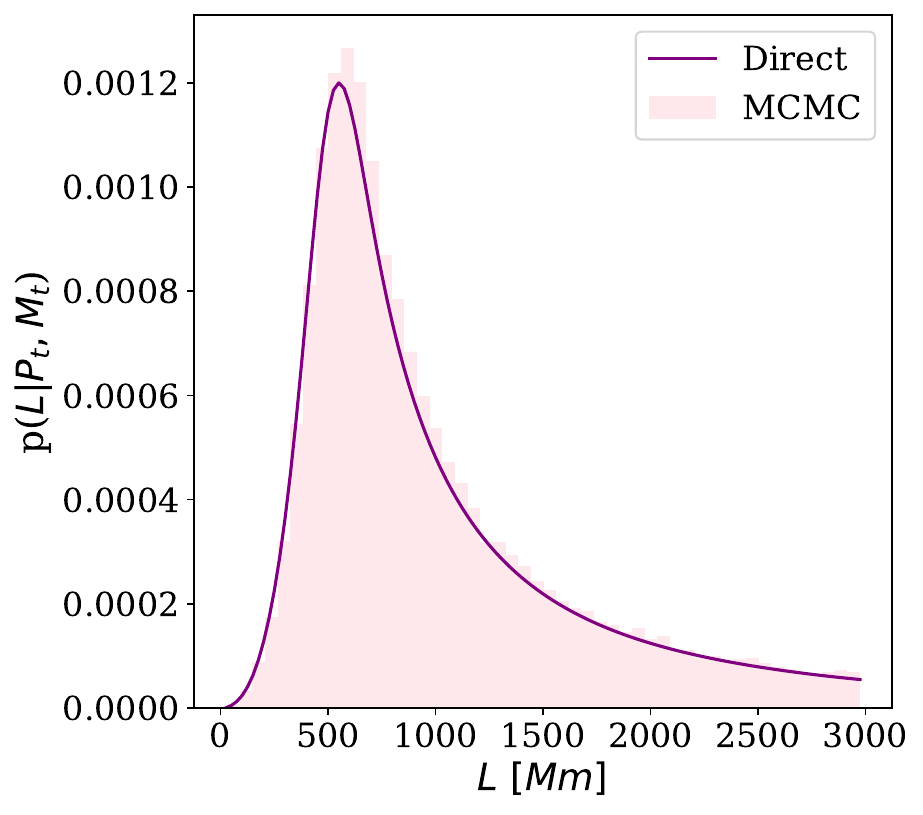}
    \caption{Joint and marginal posterior distributions of flux-tube parameters. The top-left panel shows the joint posterior distribution of magnetic field strength and flux-tube length, the top-right panel shows the joint distribution of density and flux-tube length, the bottom-left panel shows the marginal distribution of the magnetic field, and the bottom-right panel shows the marginal distribution of the flux-tube length, assuming uniform priors for all parameters and using the observational constraints of \cite{2020AA...633A..12M}. The prior on the magnetic field is taken from the posterior obtained in Sect.~\ref{sec:longitudinal_oscillations} using the uniform prior ($B_u$). Purple curves denote marginal posterior distributions obtained via Bayesian inference, while pink histograms correspond to results from the \textit{emcee} MCMC algorithm with the same dimensionality, number of walkers, steps, and burn-in phase as in Fig.~\ref{figure:longitudinal_general}. The bottom panels are not normalised~\ref{figure:longitudinal_general}.}
    \label{figure:_transverse_uniform}
\end{figure*}

The density ratio $e$ can take values between 0.6 and 2, as this range corresponds to the constant kink mode frequency \citep{2005SoPh..229...79D}, and for simplicity, it is considered to be unity, assuming that the evacuated region in the magnetic flux tube and the coronal environment have the same density. For fixed values of dimensionless variables $c$ and $l$, Equation~\ref{equation:transcedental_equation} seems to provide a unique solution for $\Omega$. However, these dimensionless variables hide five-dimensional variables ($L, W$, $\rho_c$, $\rho_p$, and $v_{\rm Ac}$) and thus infinite solutions are possible. Utilizing the least positive root ($\Omega$) of equation \ref{equation:transcedental_equation} and substituting $\omega$ as $2\pi/p_t$ in its relation with $\Omega$, and rearranging it a bit to obtain the relation between $p_t$ and $L$ in terms of $\Omega$, we have
\begin{equation}
\begin{aligned}
p_t = \frac{2 \pi L}{\Omega v_{\rm Ac}}.
\end{aligned}
\label{equation:transverse_period}
\end{equation}

\begin{figure}[t]
    \centering
    \includegraphics[width=0.45\textwidth]{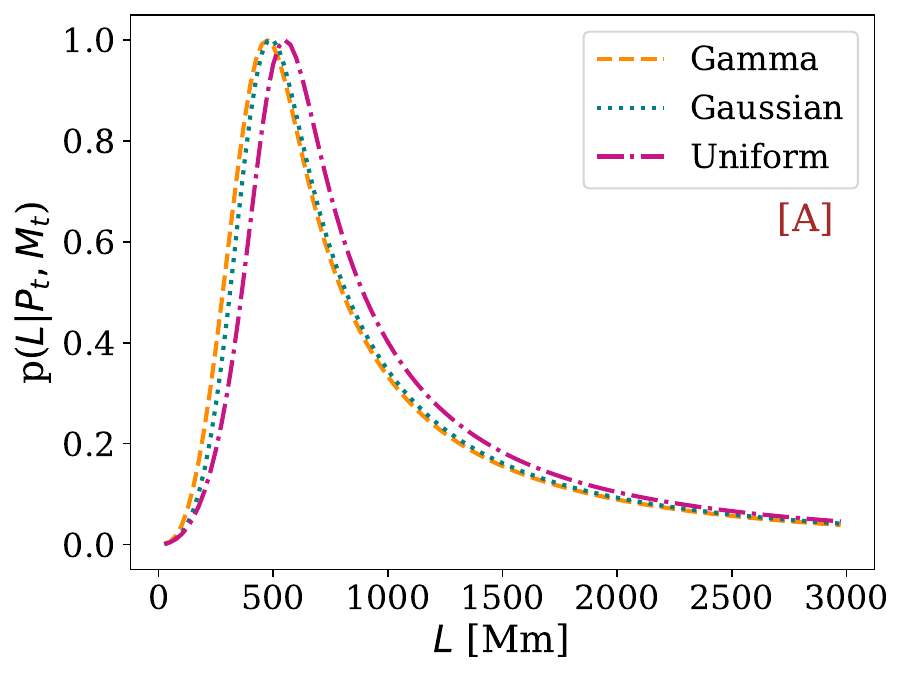}
    \includegraphics[width=0.45\textwidth]{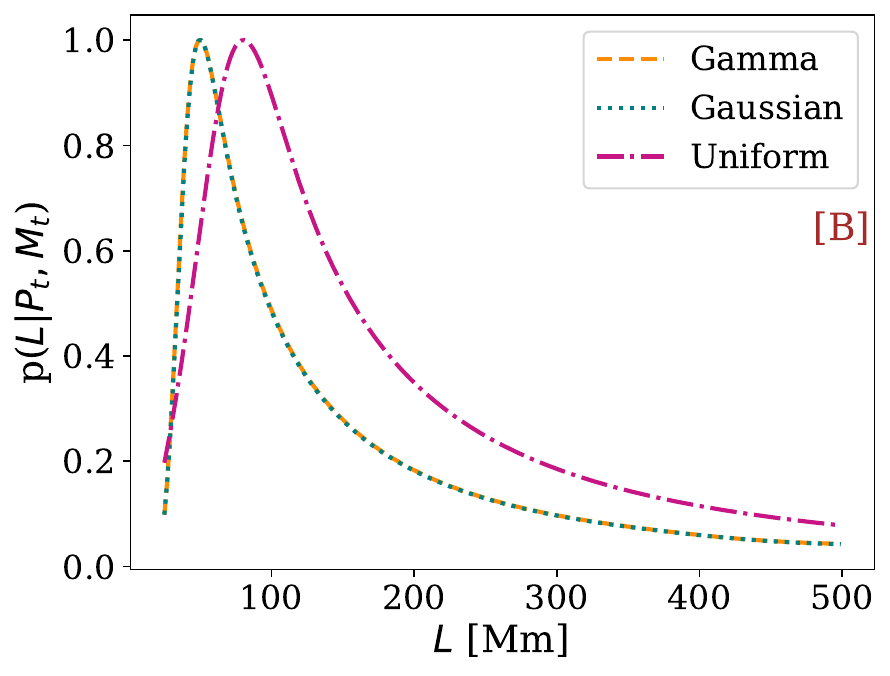}
    \caption{Marginal probability distributions of the flux-tube length. Marginal probability distributions inferred using uniform (violet dot–dashed), gamma (orange dashed), and Gaussian (teal dotted) magnetic-field priors from longitudinal oscillations in panels (A) \cite{2020AA...633A..12M} and (B) \cite{2025MNRAS.542.1308P}, with all distributions normalised to their maximum values.}
    \label{figure:marginal_trans}
\end{figure}
Further, substituting for $v_{Ac}$ in terms of the magnetic field and coronal density \(v_{Ac} = \frac{B}{\sqrt{\mu \rho_{c}}}\) in Equation \ref{equation:transverse_period}, we obtain the model for the transverse oscillations ($M_{t}$) as:
\begin{equation}
\begin{aligned}
p_t = \frac{2 \pi L \sqrt{\mu \rho_{c}}}{\Omega B}.
\end{aligned}
\label{equation:transverse_model}
\end{equation}
This equation serves as our model $M_{\rm t}$ and relates the length of the flux tube with the magnetic field strength and the period of the transverse oscillations. In what follows, we use $p_t$ and $P_t$ to denote theoretical and observational transverse oscillation periods, respectively.
Our model contains four relevant parameters, $\theta_t = \{L, B, \rho_{c}, c\}$, once the length of the prominence threads ($2W$) is fixed. In \cite{2020AA...633A..12M}, the length of the prominence thread is reported to be 1.5 Mm, and for \cite{2025MNRAS.542.1308P}, we estimated it to be 3 Mm.
For three parameters $L$, $\rho_{c}$ and $c$, we adopt uniform priors over reasonable ranges, i.e., $\mathcal{U}(L[{\rm Mm}];50,3000)$, $\mathcal{U}(\rho_c[\mathrm{kg\, m}^{-3}];10^{-14},10^{-12})$, $\mathcal{U}(c;5,200)$. The posterior of $B$ obtained from the analysis of longitudinal oscillations is now used as prior in $M_t$, thereby we have three different prior distributions for $B$ (distribution of $B_u$, $B_G$, and $B_\gamma$).
\begin{table*}
    \centering
    \caption{Transverse oscillation parameters and inferred flux-tube properties: References, transverse oscillation period ($P_t$), prominence flux-tube half length inferred using uniform ($L_u$), Gaussian ($L_G$), and gamma ($L_\gamma$) magnetic-field priors, radius of curvature ($r$) of the magnetic dip, and the associated flux-rope twist number ($\phi$), with uncertainties quoted at the 68\% credible interval.}
    \label{table:transverse_parameters}
    \begin{tabular}{ccccccccc}
    \hline
    Reference&$P_t$ [min]&$L_u$ [Mm]&$L_G$ [Mm]&$L_\gamma$ [Mm]&$r$ [Mm]&$\phi_u$&$\phi_G$&$\phi_g$\\
    \hline
    (1) &17&550$^{+575}_{-225}$&500$^{+575}_{-225}$&475$^{+550}_{-225}$&159&1.1 &0.9 &1.0\\
    (2) &16.6&69$^{+114}_{-40}$&41$^{+112}_{-20}$&43$^{+104}_{-18}$&8&2.6&1.6 &1.6\\
    \hline
    \end{tabular}
    \tablebib{(1)~\citet{2020AA...633A..12M}; (2) \citet{2025MNRAS.542.1308P}.}
\end{table*}

To construct the likelihood function, the smallest positive root of Equation \ref{equation:transcedental_equation} is used to calculate the modeled periods ($p_t$), which are then compared with the observed periods ($P_t$) of the transverse oscillations. This likelihood function, along with the prior distribution of four parameters, yields a 4D posterior probability distribution. From this 4D posterior probability distribution, the joint probability distributions for magnetic field strength ($B$) and flux tube half length ($L$), ($p(B, L|P_t, M_t)$), and for the flux tube half length ($L$) and coronal density ($\rho_{c}$), ($p(L, \rho_c|P_t, M_t)$) are obtained.

A particular example result is shown in the upper two panels of Figure \ref{figure:_transverse_uniform} for the oscillations observed by \cite{2020AA...633A..12M}. 
Additionally, the uncertainty in the periodicity of the transverse oscillations is not reported, and thus, 10$\%$ uncertainty is assumed to be associated with this parameter. 
The joint distributions indicate that both the magnetic field strength and the half length of the flux tube can be well-constrained. On the other hand, the coronal density cannot be properly inferred with the available information. The 2D joint probability distribution ($p(B, L| P_t, M_t$) is further marginalised to obtain the probable values of $B$ and $L$, which are shown in the lower panels of Figure \ref{figure:_transverse_uniform}. Both posteriors are well-constrained, although the marginal posterior for the half length of the flux tube displays a truncated long tail in the upper limit of our prior assumption. The direct integration results are validated by comparison to MCMC sampling for the posterior distribution using the \textit{emcee} algorithm. Both approaches give similar results.
The half length is around 550 Mm and thus the total length of the flux tube will be 1100 Mm, which indicates that the quiescent prominences are longer than the active region prominences.

The inference is next applied to the transverse oscillations ($P_t$) observed in \cite{2020AA...633A..12M} and \cite{2025MNRAS.542.1308P}, using the different distributions of magnetic field obtained in Section \ref{sec:longitudinal_oscillations}, to estimate the length of the prominences. The obtained results are shown in Figure~\ref{figure:marginal_trans} and the summary values reported in Table \ref{table:transverse_parameters}. Again, 10$\%$ uncertainty is assumed in the periodicity of transverse oscillations. 

All inferred posteriors have well-constrained distributions, thus the method enables to constrain the length of the magnetic flux tube from the simultaneous observation of longitudinal and transverse oscillations. The posteriors display skewed profiles with long tails towards the upper regions of $L$, leading to larger upper uncertainties  
 
Notably, the inferred lengths for both events are rather different, in spite of the similar transverse oscillation periods, because of the marked difference in the magnetic field strength values. The use of uniform priors leads to posteriors displaced towards larger values of $L$, resulting in larger maximum a posteriori estimates, following the same trend found with the inference of magnetic field strength from longitudinal oscillations. In their analysis, \cite{2020AA...633A..12M} obtain a range of $2L\in[206-713]$ Mm. Our maximum posterior estimates are contained within that range.
 
\subsection{Estimation of the twist number in flux tube}\label{sec:twist_number}
Quiescent prominences generally have inverse polarity configurations with respect to the photospheric polarity \citep{1984A&A...131...33L}. According to this opposite polarity configuration, the prominence material lies in the twisted flux tubes \citep{1974A&A....31..189K}. This configuration has been assumed in many models to explain the formation and the eruption of the prominences \citep{1989ApJ...343..971V, 2020A&A...637A..75L, 2023SoPh..298...35Z}. Thus, assuming the radius of the flux rope to be $r$, and given we have the length of the prominence, we can estimate the number of twists ($\phi$) of the flux rope where \(\phi = \frac{L}{\pi r}\). Here $r$ is calculated from the periodicity of the longitudinal oscillations using Equation \ref{equation:angular_freq}. To calculate $\phi$, we considered the most probable length obtained for each distribution for each study, and the calculated twists are reported in Table \ref{table:transverse_parameters}. The number of twists in the flux rope reported in \cite{2020AA...633A..12M} is also less than one, which matches our result using Bayesian inversion. As mentioned earlier, the length of the prominence estimated using a uniform prior of the magnetic field is larger than for the other priors, and thus, the number of twists obtained for these priors also follows the same trend. Additionally, the comparatively higher twist number obtained for \citep{2025MNRAS.542.1308P} is likely attributed to the assumption that the observed longitudinal and transverse oscillations occurred within the same prominence threads, although this was not explicitly stated in their study.

\section{Conclusions}\label{sec:Summary and conclusions}
In this study, we employed the Bayesian inversion technique to infer the physical parameters, such as the magnetic field strength and the length of prominence, by analysing the simultaneous longitudinal and transverse oscillations of the same prominences. We applied Bayesian prominence seismology to both wave modes, incorporating three prior distributions -- uniform, gamma, and Gaussian. From longitudinal oscillations, we estimated the probable values of the magnetic field strength, which are used as priors to estimate posteriors for the length of the magnetic flux tube holding prominence plasma, using transverse oscillations. Considering the flux rope model of the prominence magnetic structure, the number of turns in the associated flux tube is also estimated. 

Since no single theoretical model currently exists that can simultaneously describe both longitudinal and transverse oscillations within the same prominence, we adopted two well-established models for our analysis. The pendulum model was employed to interpret the longitudinal oscillations, while the model proposed by \citet{2005SoPh..229...79D} was used for the transverse oscillations. Although these models differ in certain aspects, they share fundamental similarities. In the pendulum model, the prominence mass is assumed to be concentrated near the magnetic dip rather than filling the entire flux tube — an assumption consistent with the \citet{2005SoPh..229...79D} model as well. Therefore, the Bayesian approach adopted in this study remains valid under the assumption that both models appropriately represent different wave modes within the same magnetic flux tube.

In contrast to analytical seismological estimates, which provide either single-values estimates for the minimum magnetic field strength \citep{2016SoPh..291.3303P,2020AA...633A..12M,2021ApJ...923...74D,2023MNRAS.520.3080T,2025MNRAS.542.1308P} or possible ranges of variation for the length of the flux tube \citep{2020AA...633A..12M}, the Bayesian framework provides full probability distributions, conditional on the assumed models and observed data. It makes use of all the available information in a consistent manner and helps to consistently propagate uncertainty from observations to inferred parameters. Our direct integration results were validated by MCMC sampling of the posteriors. 

The simplicity of the pendulum model, makes feasible the inference of the magnetic field strength with an accuracy that depends on our knowledge about the electron number density.  A shortcoming is that \cite{2022A&A...660A..54L} have estimated that, for periods above ~50 min, there are some differences between the predictions of the pendulum model and those from models that consider nonuniform gravity and noncircular dips.

Obtaining information about the length of the magnetic flux tube has traditionally been more difficult. The most direct method involves estimating the length simply from the imaging observations of filaments. This approach has been considered in numerous studies \citep{2007A&A...471..295V, 2014ApJ...795..130S}. \cite{2007Sci...318.1577O} from the wavelength of transverse thread oscillations in an active region prominence observed with Hinode, estimated a minimum length of the flux tube of 125 Mm. \cite{2019A&A...622A..88M} applied Bayesian inference to the same Hinode observations and obtained posteriors characterised by long and high tails. Their analysis suggested that the length of the flux tube can vary between 20 and 100 Mm. 

The finest approach considered to estimate the length of the flux tube is by observing both the simultaneous transverse and longitudinal oscillations in the same prominence and following the method considered by \cite{2020AA...633A..12M}. In that study, a number of assumptions on possible ranges of variation for the magnetic field strength, density, and a fixed value of the coronal Alfv\'en speed led to a range of possible values for $L$. Our study is the first obtaining full probability distributions for $L$ with uncertainty propagated from observations to inferred values. The distributions are well-constrained, but the uncertainties are large, and future efforts should improve the accuracy of the inferences.

Since there are no direct measurements of the length of the field lines, theoretical and numerical models \citep{2020A&A...637A..75L,2021A&A...654A.145L} depend on assumptions about the field lines being a few tens to a few hundred Mm long.  
From our study, it is clear that the length of the flux tubes holding the quiescent prominences can be very large (from 100 Mm to 1000 Mm). 
Future observations capturing simultaneous longitudinal and transverse oscillations within the same prominence threads could further enable the application of this method to reliably estimate both the length and twist number of the supporting magnetic flux tubes and provide this information to more advanced numerical models. 

\begin{acknowledgements}
      \textit{The python code employed for the MCMC sampling of the posteriors in this study makes use of the {\em emcee} algorithm (Foreman-Mackey et al. 2013) and was developed upon an earlier version created by M. Montes-Sol\'{\i}s. U.B. and V.P. express gratitude to ARIES for providing computational resources. I.A. acknowledges support from project PID2024-156538NB-I00 funded by MCIN/AEI/10.13039/501100011033 and by “ERDF A way of making Europe”. Finally, the authors would like to thank the referee for their valuable suggestions.
}
\end{acknowledgements}

\bibliographystyle{./bibtex/aa}
\bibliography{./bibtex/sample701.bib}
\end{document}